# Zero-field dissipationless chiral edge transport and the nature of dissipation in the quantum anomalous Hall state

Cui-Zu Chang[1]*, Weiwei Zhao[2], Duk Y. Kim[2], Peng Wei[1,3], J. K. Jain[2], Chaoxing Liu[2], Moses H. W. Chan[2]*, and Jagadeesh S. Moodera[1,3]*

[1]Francis Bitter Magnet Lab, Massachusetts Institute of Technology, Cambridge, MA 02139, USA

[2]The Center for Nanoscale Science and Department of Physics, The Pennsylvania State University, University Park, PA 16802-6300, USA

[3]Department of Physics, Massachusetts Institute of Technology, Cambridge, MA 02139, USA

**The quantum anomalous Hall (QAH) effect is predicted to possess, at zero magnetic field, chiral edge channels that conduct spin polarized current without dissipation. While edge channels have been observed in previous experimental studies of the QAH effect, their dissipationless nature at a zero magnetic field has not been convincingly demonstrated. By a comprehensive experimental study of the gate and temperature dependences of local and nonlocal magnetoresistance, we unambiguously establish the dissipationless edge transport. By studying the onset of dissipation, we also identify the origin of dissipative channels and clarify the surprising observation that the critical temperature of the QAH effect is two orders of magnitude smaller than the Curie temperature of ferromagnetism.**

Dissipationless edge transport in quantum Hall (QH) effect has resisted technological applications due to the requirements of high magnetic fields and low temperatures [1]. At least one of these two can be circumvented in the so-called quantum anomalous Hall (QAH) state in a ferromagnetic topological insulator (TI), which occurs from the combination of a topologically non-trivial inverted band structure and an intrinsic spontaneous magnetization (*M*). The most striking property of QAH state is the presence, at a zero magnetic field, of spin-polarized chiral edge channels that carry current without any dissipation whatsoever [2-4]. In the previously reported QAH state in the Cr-doped TI system [5-8], a sizable longitudinal resistance was observed (at zero magnetic field), possibly as a result of the residual nonchiral dissipative channels [9], thus hampering a direct observation of the dissipationless nature of chiral edge states in nonlocal measurements [6].

In our previous work on V-doped $(Bi,Sb)_2Te_3$, we have observed a robust zero-field quantized Hall plateau accompanied by a negligible longitudinal resistance [10]. However, it had not been unambiguously identified that the dissipationless transport was occurring via chiral edge channels. Furthermore, the physical origin of dissipation has not been clarified. The identification of dissipative channels will be essential in attempts to increase the critical temperature of the QAH state, which is almost two orders of magnitude smaller than the Curie temperature of ferromagnetism in magnetically doped $(Bi,Sb)_2Te_3$ [5-7,10]. To address these issues, we turn to nonlocal transport measurements, which provided definitive experimental evidence for the existence of chiral edge channels in the ordinary QH state at high magnetic fields [1] and also for helical edge channels in the quantum spin Hall (QSH) state at a zero magnetic field [11]. In this *Letter*, we combine two-, three- and four-terminal local and nonlocal measurements to extract information on different conducting channels and directly reveal zero-field dissipationless nature of chiral edge modes and the onset of dissipative channels for the QAH state in V-doped $(Bi,Sb)_2Te_3$.

The four quintuple layers (QL) V-doped $(Bi,Sb)_2Te_3$ films studied here were grown by molecular beam epitaxy (MBE). The transport studies were done using a dilution refrigerator (Leiden Cryogenics, 10mK, 9T) with the excitation current flowing in the film plane and the magnetic field applied perpendicular to the plane. Six-terminal Hall bridges with bottom-gate electrodes were formed in order to investigate the transport mechanism in detail [10].

We start from local transport measurements for different terminals. **Figure 1(a)** shows the magnetic field ($\mu_0 H$) dependence of the two-terminal resistances, measured at $T$=25mK, using a bottom gate bias $V_g=V_g^0$=+7V to reach the charge neutral point. The two-terminal resistances $\rho_{12,12}$, $\rho_{13,13}$ and $\rho_{14,14}$ show exactly $h/e^2$ quantized values when the magnetization $M$ is well defined. (Note that the first two subscripts of the resistivity refer to the current leads and the last two to the voltage leads.) However, local two-terminal resistance measurement cannot reveal the chirality of the edge transport explicitly.

To probe the detailed flow of the edge channels, we have performed local three-terminal resistance measurements at $T$=25mK with $V_g=V_g^0$, as shown in **Figs. 1(d)** to **1(e)**. The resistances $\rho_{14,13}$ and $\rho_{14,12}$ (**Fig.1(d)**) display similar loops that are asymmetric At $\mu_0 H$=0, the $\rho_{14,13}$ and $\rho_{14,12}$ values depend on the magnetization $M$ orientation, with a value of $h/e^2$ for $M$>0 and 0 for $M$<0, respectively. The resistances $\rho_{14,16}$ and $\rho_{14,15}$ display loops that are mirror symmetric to $\rho_{14,12}$ and $\rho_{14,13}$ by $\mu_0 H \rightarrow -\mu_0 H$, as shown in **Fig. 1(e)**. At $\mu_0 H$=0, $\rho_{14,16}$ and $\rho_{14,15}$ are 0 for $M$>0 and $h/e^2$ for $M$<0. The asymmetric loops of three-terminal resistance and the relation between $\rho_{14,16}$ ($\rho_{14,15}$) and $\rho_{14,12}$ ($\rho_{14,13}$) are direct manifestation of the chirality of edge transport, which can be understood from the Landauer-Buttiker formalism [12, 13]. In the QAH regime for $M$<0, since chiral edge modes propagate anticlockwise (1 $\rightarrow$ 2 $\rightarrow$ 3 ... ), the transmission coefficients, denoted as $T_{ij}$ from electrode $j$ to $i$, are non-zero only for $T_{21}=T_{32}=T_{43}$=1. When current flows from electrodes 1 to 4 (**Fig. 1(c)**), the voltage distributions are $V_2=V_3=V_1=(h/e^2)I$ and $V_6=V_5=V_4$=0, where $V_i$ denotes the voltage at the electrode $i$. Thus, the corresponding resistance is $\rho_{14,13}=\rho_{14,12}=0$ and $\rho_{14,14}=\rho_{14,16}=\rho_{14,15}=h/e^2$. Likewise for $M>0$, chiral edge modes travel clockwise and the nonzero transmission matrix elements are $T_{61}=T_{56}=T_{45}=1$, leading to the resistance $\rho_{14,14}=\rho_{14,13}=\rho_{14,12}=h/e^2$ and $\rho_{14,16}=\rho_{14,15}=0$. The electrical potential distributions for $M<0$ and $M>0$ (see **Fig. S9**) can also be calculated using the conformal-mapping technique [17]. Hence at $\mu_0H=0$, the two-terminal resistance $\rho_{14,14}$ is always quantized at $h/e^2$ regardless of the $M$ orientation. In contrast, the three-terminal local resistances $\rho_{14,13}$ (or $\rho_{14,12}$) and $\rho_{14,15}$ (or $\rho_{14,16}$) depend on the $M$ orientation, as seen in **Figs. 1(d)** and **1(e)**.

Both two- and three-terminal resistances deviate from the quantized value near the coercive fields ($H_c$) giving rise to two sharp resistance peaks, as shown in **Figs. 1(a), 1(d) and 1(e)**. This corresponds to plateau-to-plateau transition (PPT) region [10]. The heights of the resistance peaks for different terminal measurements are different. We plot the highest resistance values of $\rho_{12,12}$, $\rho_{13,13}$ and $\rho_{14,14}$ as a function of the distance between these terminals, as shown in **Fig. 1(b)**, and find a linear dependence between resistance and distance. This suggests that in the PPT region, the transport is occurring through the bulk of the system [14]. In contrast, the two-terminal resistance of any pair of electrodes is always $h/e^2$ independent of the length in the QAH regime, which is further consolidated by various quantized rational numbers for different interconnections among the electrodes at the periphery of the six-terminal Hall bridge (see **Figs. S1 and S2**). This $h/e^2$ quantized two-terminal resistance is similar to that in QH effect of a two-dimensional electron gas (2DEG) [15,16], with the important distinction that the ballistic edge channels of the QAH state survive without an external magnetic field. Furthermore, the ballistic transport with $h/e^2$ quantized resistance indicates the propagation of single spin species, in contrast to the QH state for which the spin polarization of the edge channels requires Zeeman coupling to an externally (usually large) applied magnetic field [1].

By comparing local measurement with non-local measurement, we can also reveal distinct behaviors for ballistic chiral edge transport and diffusive bulk transport. In **Fig. 2(a)**, the current was passed through electrodes 3 and 1, while the three-terminal local and nonlocal resistances $\rho_{31,32}$, $\rho_{31,34}$, $\rho_{31,35}$, and $\rho_{31,36}$ were measured at $T$=25mK with $V_g=V_g^0$. The $\mu_0H$ dependence of $\rho_{31,32}$ is also symmetric to those of $\rho_{31,34}$, $\rho_{31,35}$, and $\rho_{31,36}$ (**Fig. 2(b)**)) by $\mu_0H \rightarrow -\mu_0H$, thanks to the chirality of the edge current. An interesting feature is that $\rho_{31,32}$ and $\rho_{31,36}$ show dramatic peaks in the PPT region with a value of several $h/e^2$, while $\rho_{31,34}$ and $\rho_{31,35}$ vary between $h/e^2$ and 0 smoothly without any peaks. The peaks in $\rho_{31,32}$ and $\rho_{31,36}$ are induced by contributions from the local diffusive longitudinal resistance through the bulk. Since no local longitudinal resistances were picked up, no peaks appeared in $\rho_{31,34}$ and $\rho_{31,35}$.

In the above, we have shown that by carefully tuning the $V_g$, we can achieve, at the lowest temperatures, purely dissipationless chiral edge transport, and the bulk transport only occurs in the PPT region. Next we will explore the origin of dissipation at a zero magnetic field when temperature is increased. We focus on the four-terminal non-local measurement configuration where electrodes 2 and 6 were designated as the current electrodes while electrodes 3 and 5 were used as the voltage probes (**Fig. 3(b)**). Other cases are presented in [14]. In **Figs. 3(a)** and **4(h)**, at the lowest temperature $T$=25mK, $\rho_{26,35}$ is 0 (in the non-PPT region) regardless of the direction of *M*, consistent with the picture of a pure chiral edge transport [18]. With increasing temperature, $\rho_{26,35}$ exhibits a hysteresis loop with a decreasing $H_c$, as shown in **Fig. 3(a)**. The observation of hysteresis, *i.e.* high and low nonlocal resistance states at a higher temperature, indicates the appearance of other dissipative channels besides chiral edge modes. **Figure 3(d)** shows the zero-field non-local signals as a function of temperature, which increase rapidly with temperature up to 2K, accompanied by an increase of longitudinal resistance $\rho_{14,23}$ and a decrease of Hall resistance $\rho_{14,35}$, as shown in **Fig. 3(c)**. At the same time, the resistance peaks in the PPT region decay rapidly (**Fig. 3(e)**). The different temperature dependences between zero field non-local signals and the resistance peak in the PPT region suggest that they should have different origins. In Fig. 1(b), we have confirmed that the resistance peak in the PPT region follows a linear dependence on distance, thus coming from the bulk carriers. This linearity and the different temperature dependences indicate that the bulk carriers are not responsible for nonlocal signals. A possible explanation comes from the existence of *non-chiral* edge channels, which have been invoked previously to explain a similar hysteresis loop of non-local voltage in Cr doped $(Bi,Sb)_2Te_3$ [6]. In that experiment, however, the hysteresis loop was observed at the lowest temperature, indicating the existence of gapless quasihelical edge modes [9]. For our system, the observation of zero longitudinal and non-local resistances at the lowest temperature (with $V_g=V_g^0$) rules out gapless modes. Nevertheless, gapped nonchiral edge modes are possible, and, as we argue below, plausible. These non-chiral edge modes originate from two dimensional Dirac surface states on the side surfaces, which are quantized into one dimensional edge modes due to the confinement of finite thickness [9,19].

Our physical picture that the hysteresis loop of non-local signals comes from the coexistence of chiral and non-chiral edge modes finds strong support from the $V_g$ dependence of local and non-local measurements. **Figs. 4(a)** to **4(e)** show longitudinal sheet resistance $\rho_{14,23}$ and Hall resistance $\rho_{14,35}$ at different $V_g$ and the corresponding non-local resistance $\rho_{26,35}$ are shown in **Figs. 4(f)** to **4(j)**. A pronounced asymmetry between $V_g>V_g^0$ and $V_g<V_g^0$ is observed. At $T$=25mK with $V_g=V_g^0$ when the Fermi energy is in the excitation gap, $\rho_{14,35}$ is fully quantized and $\rho_{14,23}$ simultaneously vanishes. The value of non-local resistance $\rho_{26,35}$ is always 0 except at the PPT region (see **Figs. 4(c)** and **4(h)**). For $V_g>V_g^0$, $\rho_{14,23}$ and $\rho_{14,35}$ have almost no significant change (**Figs. 4(d)** and **4(e)**), while a hysteresis loop appears for the non-local resistance $\rho_{26,35}$ (**Figs. 4(i)** and **4(j)**). In contrast, for $V_g<V_g^0$, a huge $\rho_{14,23}$ is observed while the non-local resistance $\rho_{26,35}$ is always close to zero (**Figs. 4(f)** and **4(g)**). The variation of $\rho_{14,23}$ and $\rho_{14,35}$, as well as non-local resistance $\rho_{26,35}$, as a function of $V_g$ is summarized in **Figs. 4(k)** and **4(l)**. The ratio between $\rho_{26,35}$ and $\rho_{14,23}$ is around $10^{-2}$ for $M>0$ in the $V_g>V_g^0$ regime, significantly larger than its value $10^{-5}$ in the $V_g<V_g^0$ regime, as shown in **Fig. 4(m)**.

The asymmetry between $V_g>V_g^0$ and $V_g<V_g^0$ can be understood from the detailed band structure of (Bi,Sb)$_2$Te$_3$ and the positioning of the gap relative to the valence and conduction bands. From the previous ARPES measurements [20,21] and the first principles calculations [22,23], it is known that the surface Dirac cones are far away from the bulk conduction band bottom and quite close to (even buried in) the valence band. For our magnetic topological insulator system, the energy spectrum is schematically shown in **Fig. 4(o)**. The green part represents the 2D bulk bands of the thin film, which originates from both the 3D bulk bands and 2D surface bands of top and bottom surfaces. The gap of 2D bulk bands should be determined by the exchange coupling between surface states and magnetization $M$. Within the 2D bulk gap, there are two types of 1D edge modes: the chiral modes and the non-chiral edge modes originating from surface states of side surfaces as discussed earlier [9,19]. The pure dissipationless chiral edge transport only occurs when the Fermi energy is tuned into the mini-gap of non-chiral edge modes, which is induced by the confinement effect of the side surface [9] and lies close to the maximum of the valence band. For $V_g>V_g^0$, the Fermi energy first cuts through the non-chiral edge modes, leading to a hysteresis loop of non-local resistance due to the coexistence of two types of edge modes. In contrast, for $V_g<V_g^0$, the Fermi energy will first encounter the top of 2D (bulk) valence bands, resulting in a large $\rho_{14,23}$ and an insignificant non-local resistance $\rho_{26,35}$. It is known that the classical longitudinal transport can also contribute to non-local effect, for which the ratio between non-local resistance $\rho_{26,35}$ and $\rho_{14,23}$ can be estimated as $\frac{\rho^{cl}_{26,35}}{\rho_{14,23}} \approx e^{-\frac{\pi L}{W}}$, where $L$ and $W$ is the length and width of the sample [9]. In our case, $\frac{L}{W} \sim 3$, so this ratio is estimated around $8 \times 10^{-5}$, which can explain the observed non-local resistance ratio ($10^{-5}$) in the $V_g < V_g^0$ regime, but not that ($10^{-2}$) in the $V_g > V_g^0$ regime. Therefore, non-local signals as well as the hysteresis loop for $V_g > V_g^0$ should be dominated by the mechanism of the coexistence of chiral and non-chiral edge modes.

The above physical picture is also consistent the temperature dependence of longitudinal and non-local resistances in this system. With increasing temperature, $\rho_{14,23}$ increases rapidly (**Fig. 3c**), indicating the existence of bulk carriers. At the same time, the observation of hysteresis loop suggests that non-chiral edge modes should also appear. According to the band dispersion in **Fig. 4(o)**, we speculate that finite temperature excites electrons from 2D valence bands to 1D non-chiral edge channels, so that both 2D bulk holes and 1D non-chiral edge electrons coexist in the system (**Fig 4(n)**). The excitation gap, as indicated in **Fig. 4(o)**, is estimated as 50μeV by fitting temperature dependence of $\rho_{14,23}$, which is consistent with the theoretical prediction [9] and discussed in detail in the supplementary material [14]. This excitation gap is expected to be much smaller than the 2D bulk gap due to magnetization [24] and is consistent with the low critical temperature for the QAH effect. Besides these two kinds of dissipative channels, one should note that other states, such as acceptor or donor states due to impurities, could also exist in the system and cause dissipation.

In summary, our measurements provide a clear and direct confirmation of dissipationless chiral edge transport in the QAH state, identify different types of dissipative channels and provide insight into why the critical temperature for the QAH effect is two orders of magnitude smaller than the Curie temperature of the ferromagnet. The identification of dissipative channels may suggest ways to increase the critical temperature of the QAH effect, which will be crucial for its use in spintronics as well as for new chiral interconnect technologies [25]. For example, one can consider even thinner film to reduce the number of non-chiral edge modes and increase the mini-gap between non-chiral modes and valence bands. Alternatively, one can also try to reduce the Bi component in the sample to lower the energy of the valence band top, so that all the edge modes can be well above the valence band.

C.Z.C and W.Z. contributed equally to this work. We thank X. Li, H. J. Zhang, M. D. Li, S. C. Zhang and D. Heiman for helpful discussions. J.S.M. and C.Z.C acknowledge support from grants NSF (DMR-1207469), ONR (N00014-13-1-0301), and the STC Center for Integrated Quantum Materials under NSF grant DMR-1231319. M.H.W.C, W.Z. and D.Y.K. acknowledge support from NSF grants DMR-1420620（Penn State MRSEC) and DMR-1103159. J.K.J. acknowledges support from grant DOE (DE-SC0005042).

**\*Correspondence to:** czchang@mit.edu(C. Z. C.); chan@phys.psu.edu (M. H. W. C.) and moodera@mit.edu (J. S. M.).

**Figures and Figure captions:**

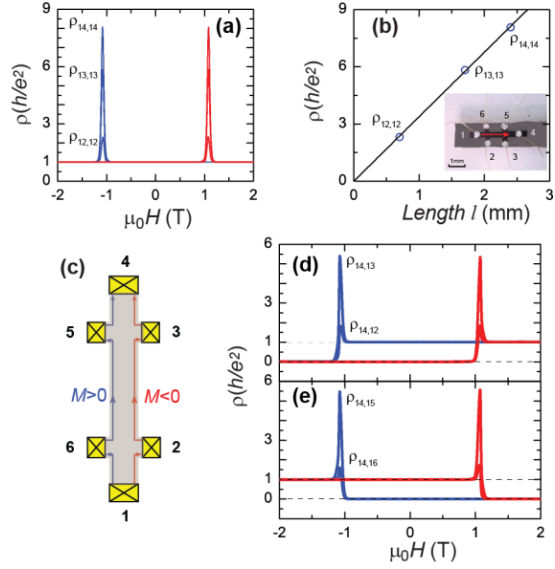

**Fig. 1.** (color online) Local two-terminal and three-terminal measurements in the QAH regime. (a) Magnetic field ($\mu_0 H$) dependence of two-terminal local resistances $\rho_{12,12}$, $\rho_{13,13}$ and $\rho_{14,14}$. (b) The peak value near the coercive field ($H_c$) of two-terminal resistances $\rho_{12,12}$, $\rho_{13,13}$ and $\rho_{14,14}$ as a function of the spacing length $l$(mm) between the voltage electrodes. The inset photograph shows the Hall bridge device. (b) Schematic layout of the device applicable for panels (d) and (e) and also for $\rho_{14,14}$ in (a). The current flows from 1 to 4. The red and blue lines indicates the chiral edge current for magnetization into ($M<0$) and out of the plane ($M>0$), respectively. (d, e) $\mu_0 H$ dependence of three-terminal resistances $\rho_{14,13}$, $\rho_{14,12}$ (d) and $\rho_{14,15}$, $\rho_{14,16}$ (e).

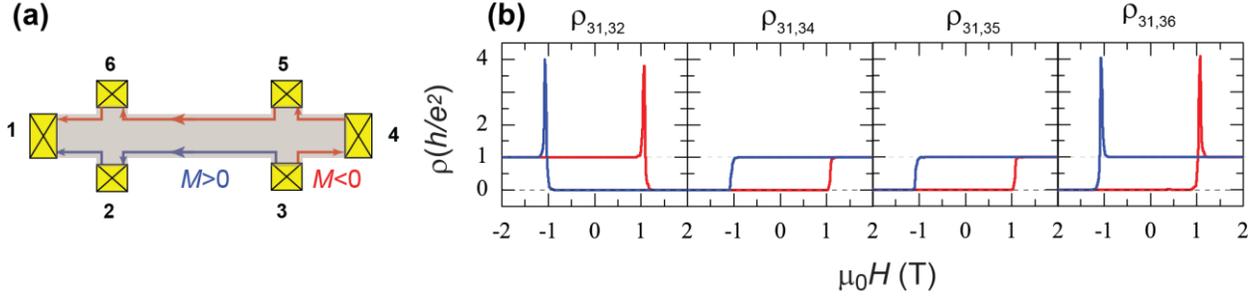

**Fig. 2.** (color online) Local and nonlocal three-terminal measurements in the QAH regime. (a) Chiral edge conduction channels when the current flows from 3 to 1. (b) $\mu_0H$ dependence of local and nonlocal three-terminal resistances $\rho_{31,32}$, $\rho_{31,34}$, $\rho_{31,35}$, and $\rho_{31,36}$ measured at $T=25$ mK with $V_g=V_g^0$.

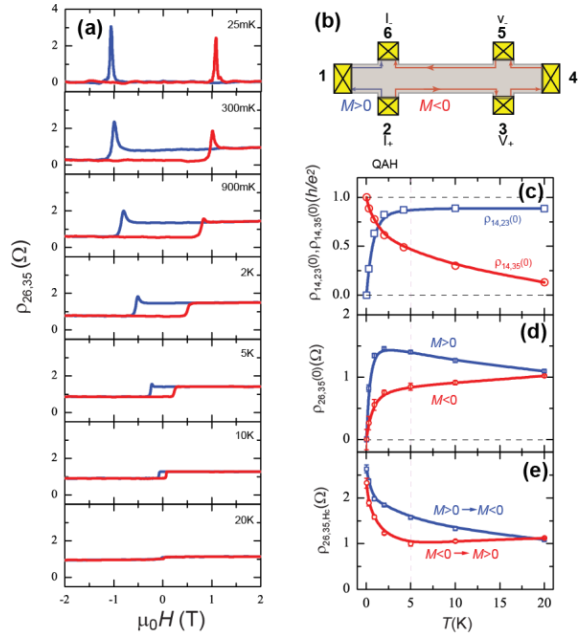

**Fig. 3.** (color online) Temperature dependence of chiral edge transport in QAH regime. (a) $\mu_0H$ dependence of the nonlocal resistance $\rho_{26,35}$ measured at $V_g=V_g^0$ from 25 mK to 20 K. (b) Chiral edge conduction channels when the current flows from 2 to 6, and the nonlocal voltage measured between 3 and 5. (c) Temperature dependence of the zero-field longitudinal sheet resistance $\rho_{14,23}(0)$ (blue curve) and Hall resistance $\rho_{14,35}(0)$ (red curve). (d) Temperature dependence of zero-field nonlocal signal $\rho_{26,35}(0)$ for $M<0$ (red curve) and $M>0$ (blue curve), respectively. (e) Temperature dependence of nonlocal resistance $\rho_{26,35}$ peaks in the PPT as seen in (a) going between two magnetization orientations.

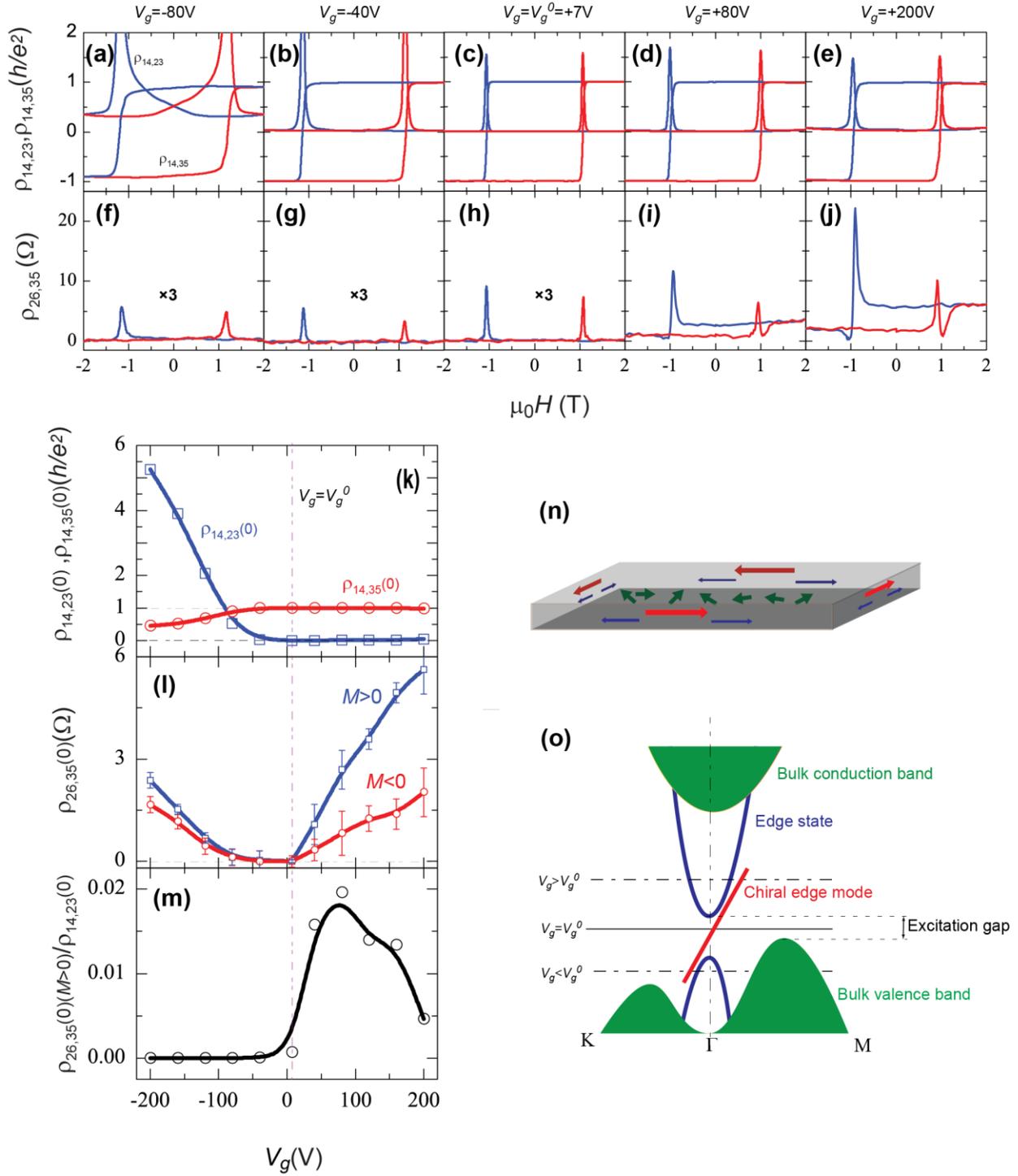

**Fig. 4.** (color online) Gate bias dependence of chiral edge transport in QAH regime. (a to j) $\mu_0 H$ dependence of longitudinal resistance $\rho_{14,23}$ and Hall resistance $\rho_{14,35}$ (a to e), as well as the nonlocal signal $\rho_{26,35}$ (f to j) at various $V_g$s. (k) $V_g$ dependence of the zero-field longitudinal sheet resistance $\rho_{14,23}(0)$ (blue curve) and Hall resistance $\rho_{14,35}(0)$ (red curve). (l) $V_g$ dependence of zero-field nonlocal signal $\rho_{26,35}(0)$ for $M<0$ (red curve) and $M>0$ (blue curve), respectively. (m) $V_g$ dependence of the ratio between non-local resistance $\rho_{26,35}(0)$ for $M>0$ and longitudinal sheet resistance $\rho_{14,23}(0)$. Note that the non-local resistance $\rho_{26,35}(0)$ for $M>0$ indicates that the dissipation channels in the voltage probe side. (n) The schematic diagrams identify three channels in the sample: dissipationless edge channels (red arrows), dissipative edge channels (blue arrows) and dissipative bulk channels (green arrows). (o) The schematic band dispersions of the sample. The horizontal dash-dotted line indicates the Fermi level position for $V_g>V_g^0$ and $V_g<V_g^0$, respectively. The excitation gap is defined as the energy difference between the bottom of the non-chiral edge mode and the maximum of bulk valence band as indicated.

# Supplementary Materials for

# Zero-field dissipationless chiral edge transport and the nature of dissipation in the quantum anomalous Hall state

Cui-Zu Chang[1]*, Weiwei Zhao[2], Duk Y. Kim[2], Peng Wei[1], J. K. Jain[2], Chaoxing Liu[2], Moses H. W. Chan[2]*, and Jagadeesh S. Moodera[1,3]*

[1]Francis Bitter Magnet Lab, Massachusetts Institute of Technology, Cambridge, MA 02139, USA

[2]The Center for Nanoscale Science and Department of Physics, The Pennsylvania State University, University Park, PA 16802-6300, USA

[3]Department of Physics, Massachusetts Institute of Technology, Cambridge, MA 02139, USA

*Correspondence to: czchang@mit.edu (C. Z. C.); chan@phys.psu.edu (M. H. W. C.) and moodera@mit.edu (J. S. M.)

**This PDF file includes:**

Materials and Methods

Supplementary Text

Figures S1-S10

References

# Ⅰ. Materials and Methods

**MBE growth**. Thin film growth was performed using a custom-built ultrahigh vacuum MBE system with a base vacuum of $2\times10^{-10}$ Torr. Heat-treated insulating SrTiO$_3$(111) substrates were degassed before the growth of TI films. High-purity Bi(99.999%), Sb(99.9999%), and Te(99.9999%) were evaporated from Knudsen effusion cells, whereas the transition metal dopant V(99.995%) were evaporated by a e-gun crucible. During the growth, the substrate was maintained at 230℃. The flux ratio of Te to Bi and Sb was set to approximately ~8 to minimize Te deficiency in the films. The concentrations of Sb, Bi, and V in the films were determined by their ratio obtained in situ during the growth using separate quartz crystal monitors and later confirmed ex situ by inductively coupled plasma atomic emission spectroscopy (ICP-AES). The growth rate for the films was approximately 0.2 quintuple layers per minute. Epitaxial growth was monitored by in situ reflection high energy electron diffraction (RHEED) patterns, where the high crystal quality and the atomically flat surface were confirmed by the streaky and sharp "1×1" patterns [S1].

**Electrical transport measurements.** The transport measurements were performed ex situ on the magnetically doped TI thin films. To avoid possible surface and film degradation, a 10nm thick epitaxial Te capping layer was deposited at room temperature on top of the TI films before it was taken out of the growth chamber for transport measurements. The electrical transport studies were done using a dilution refrigerator (Leiden Cryogenics, 10mK, 9T) with the excitation current flowing in the film plane with the magnetic field applied perpendicular to the plane. The bottom gate voltage was applied using a Keithley 6430 voltage source. All the current and volt meters were calibrated by a standard resistor.

## Ⅱ. Supplementary Text

### ⅰ)Quantized two-terminal resistance in multiply-connected perimeters.

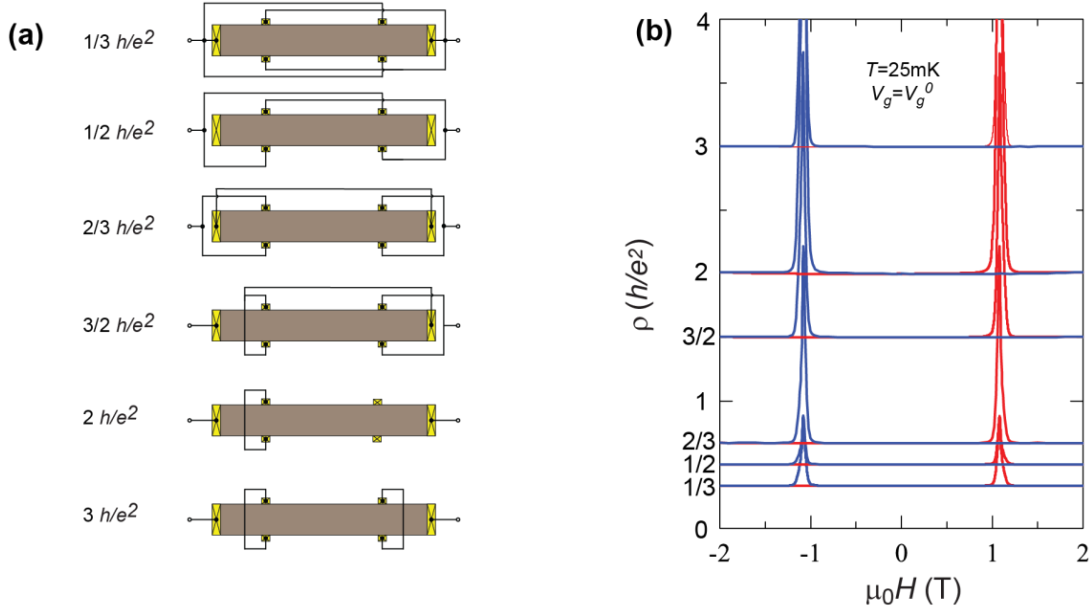

**Fig. S1. Quantized two-terminal resistance in multiply-connected perimeters.** **(a)** The perimeter interconnection configurations of the six terminals Hall bridge. The quantized values of two-terminal resistance are indicated on the left of the configurations. **(b)** $\mu_0 H$ dependence of two-terminal resistance for each configuration in **(a)**. Red and blue colors represent two directions of magnetic field sweep.

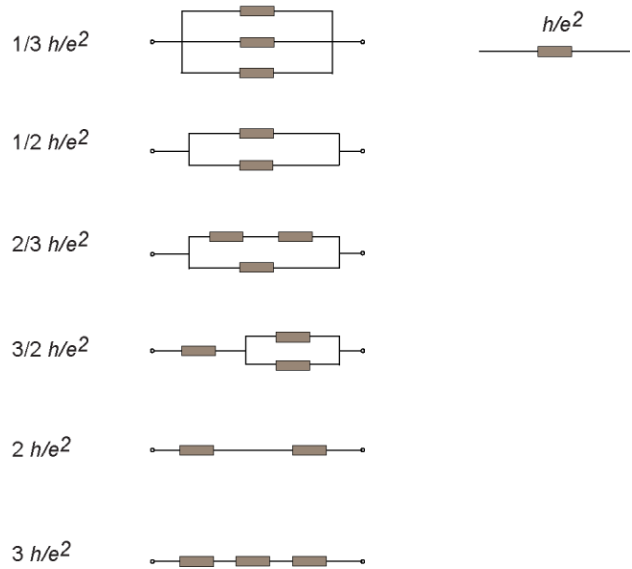

**Fig. S2.** The corresponding circuit diagrams for the configurations of **Fig. S1a**.

To further consolidate our observations of the ballistic chiral edge conduction channels, we made different interconnections among the electrodes at the periphery of the six-terminal Hall bridge [S2], as shown in **Fig. S1a**. Resistances quantized at a variety of rational fractions and multiples of $h/e^2$, such as 1/3, 1/2, 2/3, 1(see **Fig. 1a** of the main text), 3/2, 2, and 3, were obtained, as shown in **Fig. S1b**.

**Figure S2** shows the corresponding circuit diagrams for the configurations in **Fig. S1a**. In the QAH regime, for multiple connected peripheral contacts, only dissipative regions are in the vicinity of the current probes. In general, each adjacent pair of the contact must satisfy the $h/e^2$ quantized two-terminal resistances, as shown in **Fig. 1a** of the main text. So the various quantized rotationally-numbered two-terminal resistances observed in **Fig. S1b** can be easily understood by series and parallel connection of two-terminal quantized $h/e^2$ resistance, as shown in **Fig. S2**. In principle, if there are *2N* connections in the ideal QAH sample, one may easily obtain $(m/n)h/e^2$ quantized resistances at zero magnetic field by multiply connected peripheral contacts, here *m*, *n= 1, 2, ……N*. All these quantized resistances are independent of sample size, sample shape, and the electrode positions [S2].

ⅱ）**Temperature dependence of local two-, three-, four-terminal measurements**

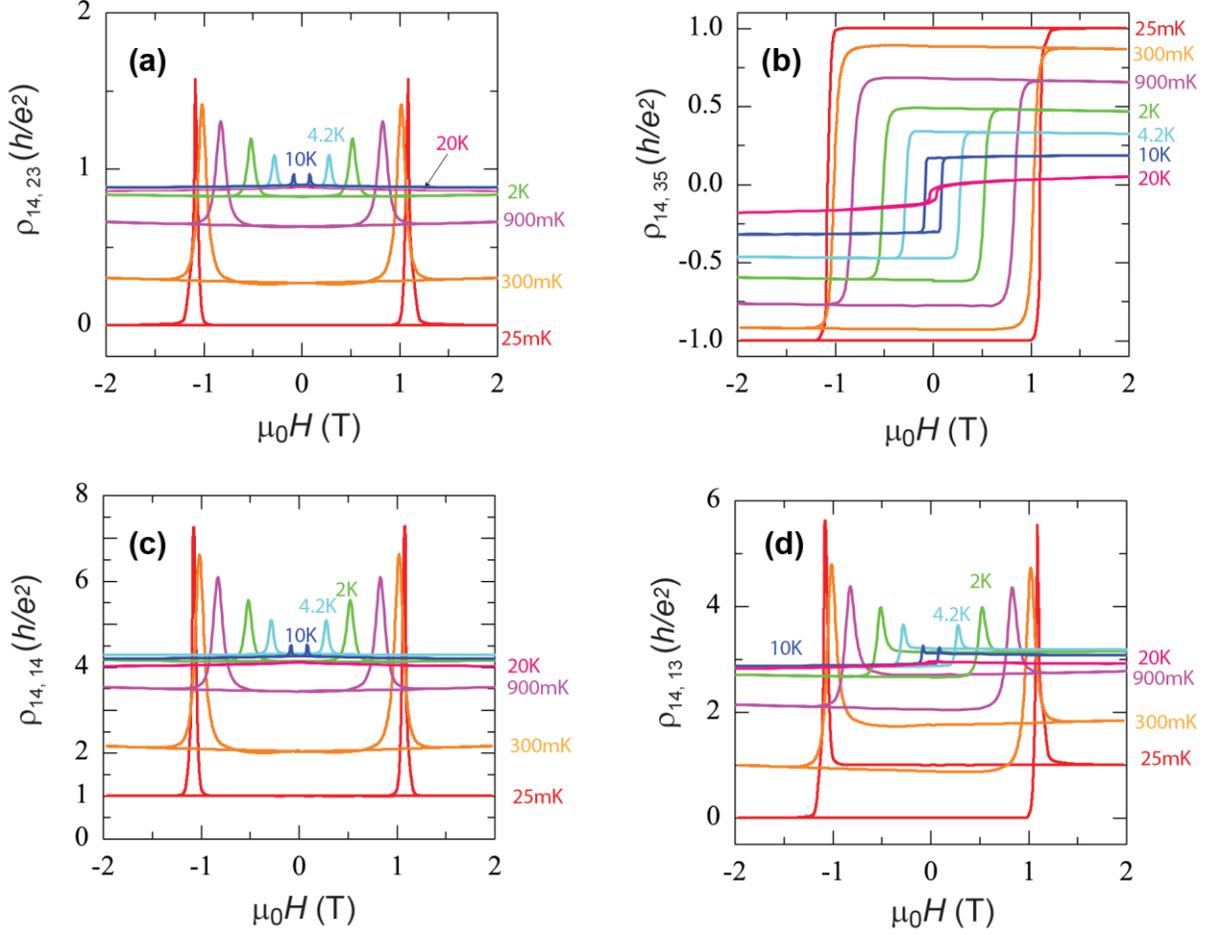

**Fig. S3. Temperature dependence of local two-, three-, four-terminal measurements at $V_g=V_g^0$.** **(a-d)** $\mu_0H$ dependence of four-terminal longitudinal sheet resistance $\rho_{14,23}$ **(a)**, Hall resistance $\rho_{14,35}$ **(b)**, two-terminal resistance $\rho_{14,14}$ **(c)**, three-terminal resistance $\rho_{14,13}$ **(d)** at various temperatures.

The $\mu_0H$ dependence of four-terminal longitudinal sheet resistance $\rho_{14,23}$ and Hall resistance $\rho_{14,35}$ at different temperature are shown in **Figs.S3a** and **S3b**. With increasing temperature, the $\rho_{14,23}$ at zero-field (labeled as $\rho_{14,23}(0)$) increases from 0 at $T=25$mK to ~$0.85h/e^2$ at $T=20$K, and the corresponding $\rho_{14,35}$ at zero-field (labeled as $\rho_{14,35}(0)$) decreases from exactly $h/e^2$ at $T=25$mK to ~$0.01 h/e^2$ at $T=20$K, due to the introduction of dissipative channels. At $T=25$mK, the system conduction is purely via chiral edge channels; 25mK<$T$<5K, the system conducts with a mixture of chiral edge channels and dissipative nonchiral edge/bulk conduction; For $T>5$K, the conduction of the system becomes dominantly dissipative nonchiral edge/bulk conduction. Another salient feature in **Fig. S3a** is that, the absolute value of the peak at $H_c$ systematically becomes smaller with increasing temperature. The large peak at the lowest temperature $T=25$mK can be understood by the QAH phenomenology. The magnetization reversal of a QAH system leads to a quantum phase transition between two QH states, in which the dissipationless chiral edge state vanishes and the system becomes an ordinary insulator with significantly enhanced resistance. This property was also reflected by the two- and three-terminal measurements at various temperatures, as shown in **Figs. S3c** and **S3d**. When only pure dissipationless chiral edge channels exist, the system can have length-independent resistances.

ⅲ）**Gate dependence of two-, three-, four-terminal measurements**

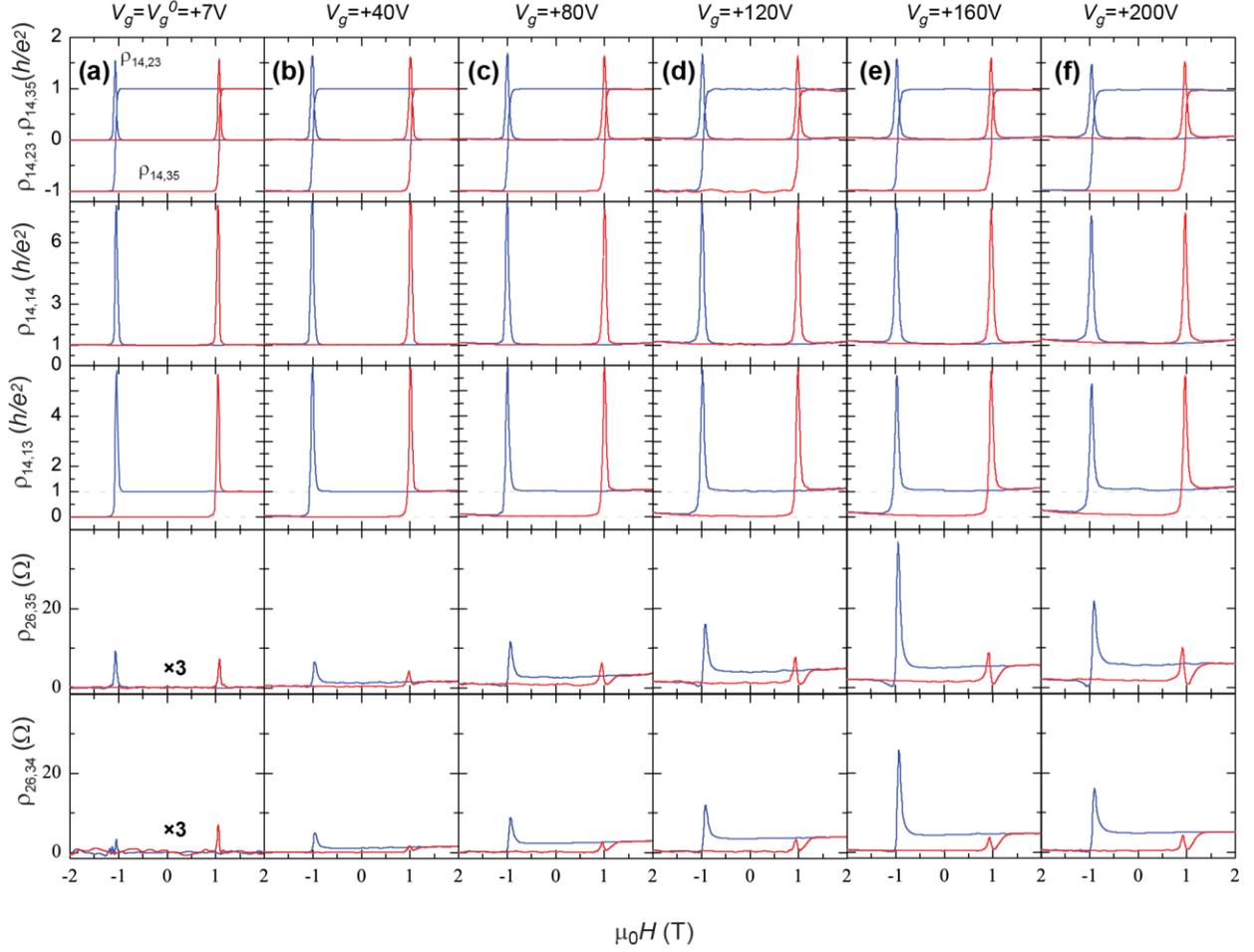

**Fig. S4. Gate dependence of two-, three-, four-terminal measurements at $T$=25mK with $V_g > V_g^0$. (a-f)** $\mu_0H$ dependence of four-terminal longitudinal sheet resistance $\rho_{14,23}$ and Hall resistance $\rho_{14,35}$, two-terminal resistance $\rho_{14,14}$, three-terminal resistance $\rho_{14,13}$, and nonlocal four-terminal resistance $\rho_{26,35}$, $\rho_{26,34}$ at various gate bias $V_g = V_g^0$=+7V **(a)**, $V_g$=+40V **(b)**, $V_g$=+80V **(c)**, $V_g$=+120V **(d)**, $V_g$=+160V **(e)**, $V_g$=+200V **(f)**. Red and blue colors represent two directions of magnetic field sweep.

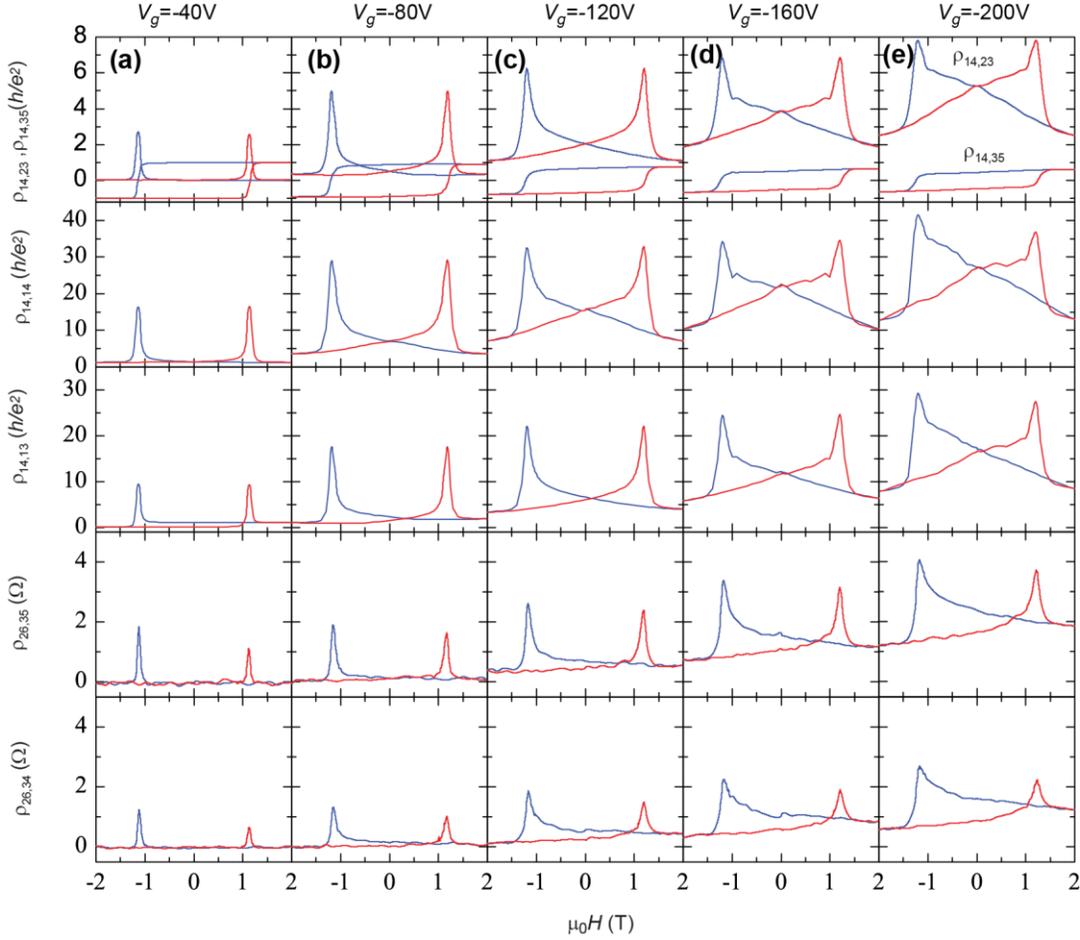

**Fig. S5. Gate dependence of two-, three-, four-terminal measurements at $T=25$mK with $V_g<V_g^0$. (a-f)** $\mu_0H$ dependence of four-terminal longitudinal sheet resistance $\rho_{14,23}$ and Hall resistance $\rho_{14,35}$, two-terminal resistance $\rho_{14,14}$, three-terminal resistance $\rho_{14,13}$, and nonlocal four-terminal resistance $\rho_{26,35}$, $\rho_{26,34}$ at various gate bias $V_g=$ -40V **(a)**, $V_g=$-80V **(b)**, $V_g=$-120V **(c)**, $V_g=$-160V **(d)**, $V_g=$-200V **(e)**. Red and blue colors represent two directions of magnetic field sweep.

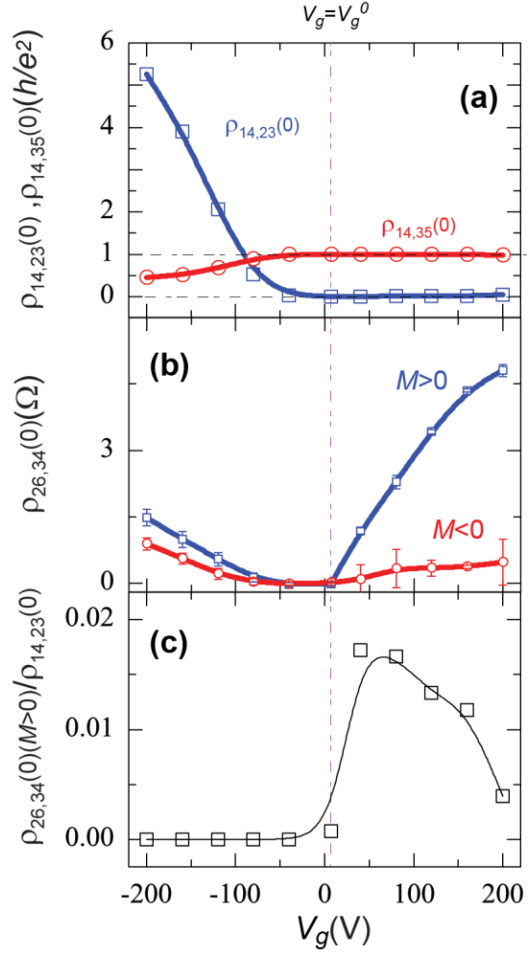

**Fig. S6. Gate dependence of the QAH effect measured at $T= 25$mK. (a)** $V_g$ dependence of zero-field longitudinal sheet resistance $\rho_{14,23}(0)$ (blue open squares) and zero-field Hall resistance $\rho_{14,35}(0)$ (red open circles). **(b)** $V_g$ dependence of zero-field nonlocal four-terminal signal $\rho_{26,34}(0)$ for negative magnetization ($M<0$) (red curve) and positive magnetization $M>0$(blue curve), respectively. **(c)** $V_g$ dependence of the ratio between $\rho_{26,34}(0)$ for $M>0$ and $\rho_{14,23}(0)$. Note that the nonlocal resistance $\rho_{26,34}(0)$ for $M>0$ indicates that the dissipationless chiral edge channel doesn't go through the $2 \rightarrow 3 \rightarrow 4 \rightarrow 5 \rightarrow 6$ side of the sample.

**Figures S4** and **S5** show the $\mu_0H$ dependence of four-terminal longitudinal sheet resistance $\rho_{14,23}$ and Hall resistance $\rho_{14,35}$, two-terminal resistance $\rho_{14,14}$, three-terminal resistance $\rho_{14,13}$, and nonlocal four-terminal resistance $\rho_{26,35}$ and $\rho_{26,34}$ at various gate bias $V_g$. A profound asymmetry between $V_g > V_g^0$ and $V_g < V_g^0$ is observed. At the lowest temperature $T=25$mK with $V_g = V_g^0$ when the Fermi energy is in the excitation gap (see **Fig. 4o** of the main text), $\rho_{14,35}$ is fully quantized and $\rho_{14,23}$ simultaneously vanishes. The value of non-local resistance $\rho_{26,35}$ is always 0 except at $H_c$. For $V_g > V_g^0$, $\rho_{14,23}(0)$ and $\rho_{14,35}(0)$ have modest change (**Fig.S4**): the $\rho_{14,23}(0)$ increases only from 0 at $V_g = V_g^0 = +7$V to ~0.05 $h/e^2$ (~1216Ω) at $V_g = +200$V, the $\rho_{14,35}(0)$ decreases from $h/e^2$ at $V_g = V_g^0$ to ~0.978 $h/e^2$ at $V_g = +200$V due to the introduction of dissipative nonchiral edge channels (**Fig. S6a**). As a result, the corresponding $\rho_{14,14}$ at zero-field (labeled as $\rho_{14,14}(0)$) increases from exactly $h/e^2$ at $V_g = V_g^0$ to ~1.05 $h/e^2$ at $V_g = +200$V, $\rho_{14,13}$ at zero-field (labeled as $\rho_{14,13}(0)$) increases from $h/e^2$ for $M > 0$ and 0 for $M < 0$ at $V_g = V_g^0$ to ~1.13 $h/e^2$ for $M > 0$ and ~0.09 $h/e^2$ for $M < 0$ at $V_g = +200$V, and the corresponding nonlocal signal $\rho_{26,35}$ and $\rho_{26,34}$ gradually exhibit hysteresis loops when $V_g > V_g^0$. In contrast, for $V_g < V_g^0$, a distinct change of $\rho_{14,23}$ and $\rho_{14,35}$ is observed while the non-local resistance $\rho_{26,35}$ and $\rho_{26,34}$ is always close to zero (**Fig.S5**). The $\rho_{14,23}(0)$ increases only from 0 at $V_g = V_g^0$ to ~5.3 $h/e^2$ at $V_g = -200$V, the $\rho_{14,35}(0)$ decreases from $h/e^2$ at $V_g = V_g^0$ to ~0.446 $h/e^2$ at $V_g = -200$V, due to the introduction of dissipative bulk valence channels (**Fig. S6a**). As a result, the corresponding $\rho_{14,14}(0)$ increases from exactly $h/e^2$ at $V_g = V_g^0$ to ~27.3 $h/e^2$ at $V_g = -200$V, $\rho_{14,13}(0)$ increases from $h/e^2$ for $M > 0$ and 0 for $M < 0$ at $V_g = V_g^0$ to ~16.7 $h/e^2$ for both $M > 0$ and $M < 0$ at $V_g = -200$V. $\rho_{26,35}$ and $\rho_{26,34}$ always have a small value close to zero and no obvious loops as $V_g > V_g^0$ appears.

The $V_g$ dependence of $\rho_{26,34}(0)$ are plotted in **Fig. S6b**. Compared with $\rho_{26,35}(0)$ (see **Fig. 4l** of the main text), the more obvious slope difference in $\rho_{26,34}(0)$ for $M < 0$ at $V_g > V_g^0$ also indicates that the diffusive nonchiral edge channels survive. The ratio between nonlocal resistance $\rho_{26,34}(0)$ for $M > 0$ and $\rho_{14,23}(0)$ is around $10^{-2}$ in the $V_g > V_g^0$ regime due to the introduction of diffusive nonchiral edge channels, however, this ratio is only around $10^{-5}$ in the $V_g<V_g^0$ regime due to the classical transport induced by the diffusive bulk channels, as shown in **Fig. S6c**. These results further support the discussion of the main text.

ⅳ）**Other four-terminal nonlocal measurements in the QAH regime**

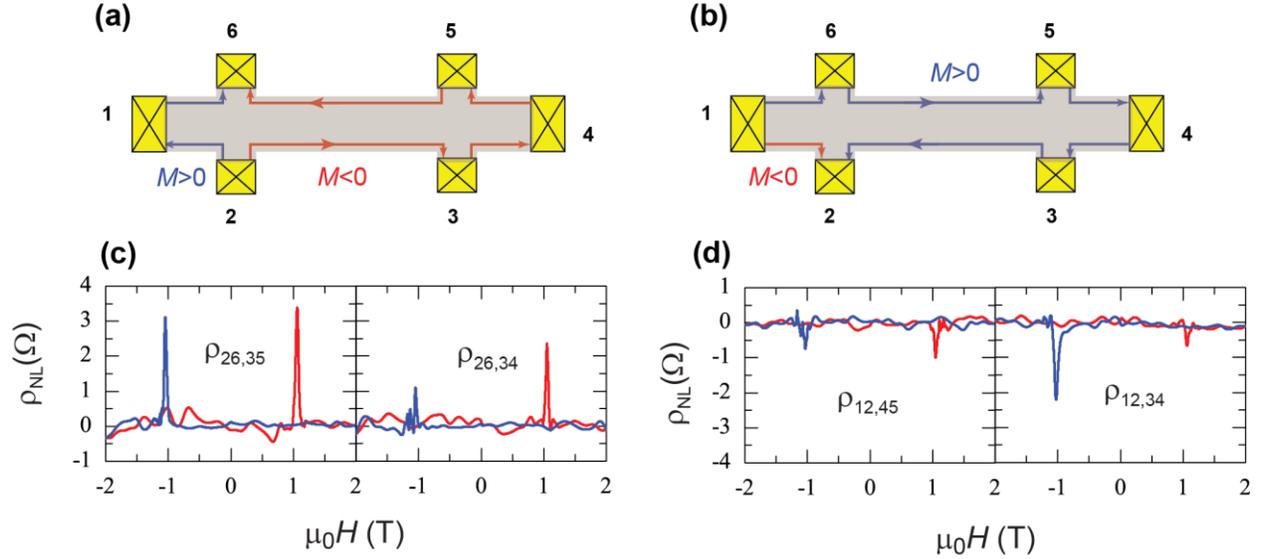

**Fig. S7. Four-terminal nonlocal measurements of the six terminals Hall bridge at $V_g=V_g^0$. (a, b)** Schematic layout of the chiral edge conduction channels when the current is applied through the electrodes 2 to 6 (**a**) and 1 to 2 (**b**). (**c**) $\mu_0H$ dependence of the nonlocal signal $\rho_{26,35}$ and $\rho_{26,34}$. (**d**) $\mu_0H$ dependence of the nonlocal signal $\rho_{12,45}$ and $\rho_{12,34}$. All the measurements were taken at $T$=25mK. Red and blue colors in (**c**) and (**d**) represent two directions of magnetic field sweep.

Apart from the four-terminal nonlocal measurement in the main text, we also investigated other four-terminal nonlocal configurations, *i.e.* $\rho_{26,35}$ and $\rho_{26,34}$ (see **Figs. S7a** and **S7c**); $\rho_{12,45}$ and $\rho_{12,34}$ (see **Figs. S7b** and **S7d**). In all cases, the values were equal to 0 in both negative and positive of magnetization (*M*) orientations at zero magnetic field, further ascertaining the truly dissipationless chiral edge channels in the QAH state. The peaks in $\rho_{26,35}$ and $\rho_{26,34}$ and the dips of $\rho_{12,45}$ and $\rho_{12,34}$ indicate the expected chirality of the edge channels in the perfect QAH regime.

ⅴ）**Other three-terminal local and nonlocal measurements in the QAH regime**

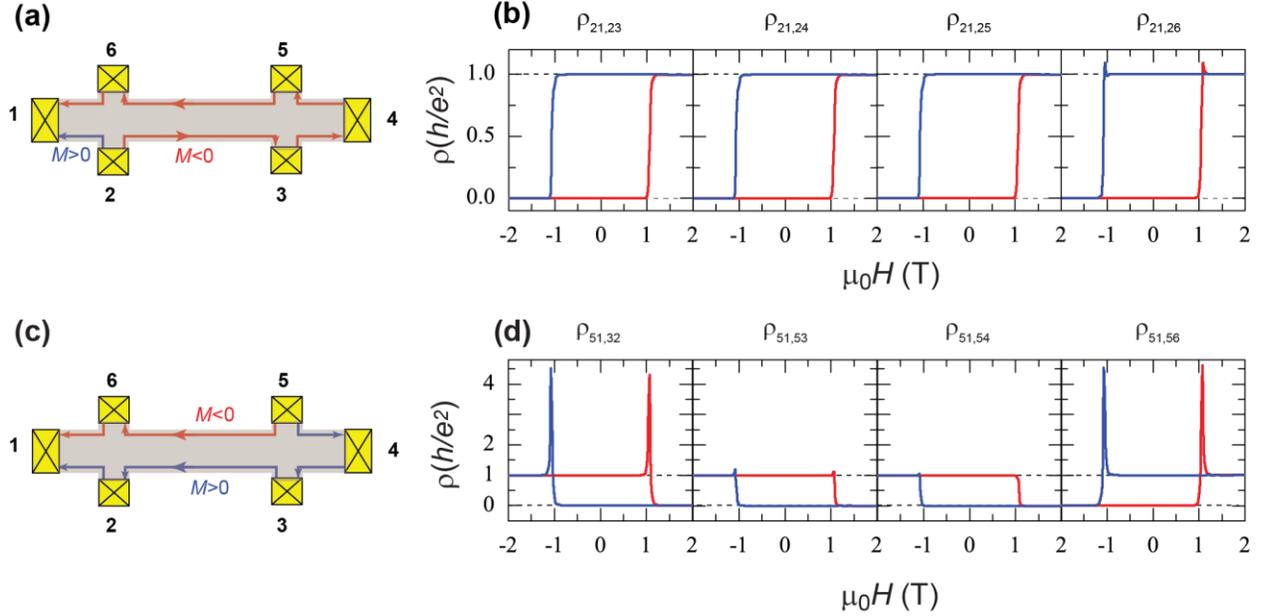

**Fig. S8. Three-terminal local and nonlocal measurements of the six terminals Hall bridge at $V_g=V_g^0$.** (a) Schematic layout of the chiral edge conduction channels when the current is applied through the electrodes 2 to 1. (b) $\mu_0H$ dependence of local and nonlocal signal $\rho_{21,23}$ and $\rho_{21,24}$. $\rho_{21,24}$ and $\rho_{21,26}$. (c) Schematic layout of the chiral edge conduction channels when the current is applied through the electrodes 5 to 1. (d) $\mu_0H$ dependence of local and nonlocal signal $\rho_{51,52}$ and $\rho_{51,53}$. $\rho_{51,54}$ and $\rho_{51,56}$. All the measurements were taken at $T=25$mK. Red and blue colors in (b) and (d) represent two directions of magnetic field sweep.

In addition to the three-terminal measurements shown in **Fig. 2** of the main text, we also investigated other three-terminal measurements. Flowing the current through electrodes 2 and 1, the nonlocal three terminal resistances $\rho_{21,23}$, $\rho_{21,24}$, $\rho_{21,25}$, and $\rho_{21,26}$ were measured, as shown in **Fig. S8a**. The $\mu_0H$ dependence of these three-terminal nonlocal resistances, measured at $T=25$mK with $V_g=V_g^0$, are shown in **Fig. S8b**. $\rho_{21,23}$, $\rho_{21,24}$, $\rho_{21,25}$ and $\rho_{21,26}$ reveal square loops, with a value of $h/e^2$ for $M>0$ and 0 for $M<0$. This is expected from the pure chiral edge transport in the Landauer-Buttiker formalism [S3,S4]. The initial conditions are $V_2=V$, $V_1=0$, $I_2=-I_1=I$, $I_3=I_4=I_5=I_6=0$. In pure chiral edge transport regime, one has $V_6=V_5=V_4=V_3=V_2=(h/e^2)I$ for $M<0$, and $V_6=V_5=V_4=V_3=V_1=0$, $V_2=(h/e^2)I$ for $M>0$. Nevertheless, $\rho_{21,26}$ shows additional features, two small peaks appear in the PPT region. In another approach, the current was passed through electrodes 5 and 1, while the three-terminal local and nonlocal resistances $\rho_{51,52}$, $\rho_{51,53}$, $\rho_{51,54}$, and $\rho_{51,56}$ were measured, as shown in **Fig. S8c**. The $\mu_0H$ dependence of local and nonlocal three-terminal resistances $\rho_{51,52}$ and $\rho_{51,53}$. $\rho_{51,54}$ and $\rho_{51,56}$, measured at $T=25$mK with $V_g=V_g^0$ are shown in **Fig. S8d**. $\rho_{51,56}$ shows the mirror symmetric loop, compared with $\rho_{51,52}$, $\rho_{51,53}$, and $\rho_{51,54}$, as a result of the chirality of the edge current. Both $\rho_{51,52}$ and $\rho_{51,56}$ show sharp peaks at the PPT region. The peaks in $\rho_{21,26}$, $\rho_{51,52}$ and $\rho_{51,56}$ are attributed to the induced contributions from the local diffusive longitudinal resistance through the bulk. The reason why no or obscure peaks appear in $\rho_{21,23}$, $\rho_{21,24}$, $\rho_{21,25}$, $\rho_{51,53}$ and $\rho_{51,54}$ could be that they do not pick up or pick up only a small fraction of the local longitudinal resistance. These experiments show that nonlocal transport of dissipationless chiral edge channels only depends on relative position between two voltage probes while the dissipative bulk transport depends on spacing between two voltage probes.

ⅵ）**The calculated electric potential distributions of six terminals Hall bridge in the QAH regime**

In order to understand the data in **Fig.1** of the main text, we calculate the electrical potential distributions for *M*<0 (**Fig. S9a**) and *M*>0 (**Fig. S9b**) when the current is passed from the electrode 1 to 4 using the conformal-mapping technique [S5]. In the QAH regime, due to the large value of Hall angle ($\alpha$), most of the electric field is concentrated at two accumulation regions at opposite corners of the sample, which are known as "hot spots" [S6, S7], as shown in **Fig. S9**. In the "hot spot" corners, the electric field exhibits a power-law singularity, and thus the electrical potential changes suddenly from 0 to *V* or *V* to 0 [S5]. From the electrical potential distributions of Hall bridge for *M*<0 (**Fig. S9a**) and *M*>0 (**Fig. S9b**), the two-terminal resistances $\rho_{14,14}$ show exactly $h/e^2$ quantized values only except in the narrow magnetic field region when the magnetization *M* of the sample is being reversed. The three-terminal resistances $\rho_{14,13}$ and $\rho_{14,12}$ values are $h/e^2$ for *M*>0 and 0 for *M*<0, whereas $\rho_{14,16}$ and $\rho_{14,15}$ are 0 for *M*>0 and $h/e^2$ for *M*<0 (see **Figs. 1d** and **1e** of the main text).

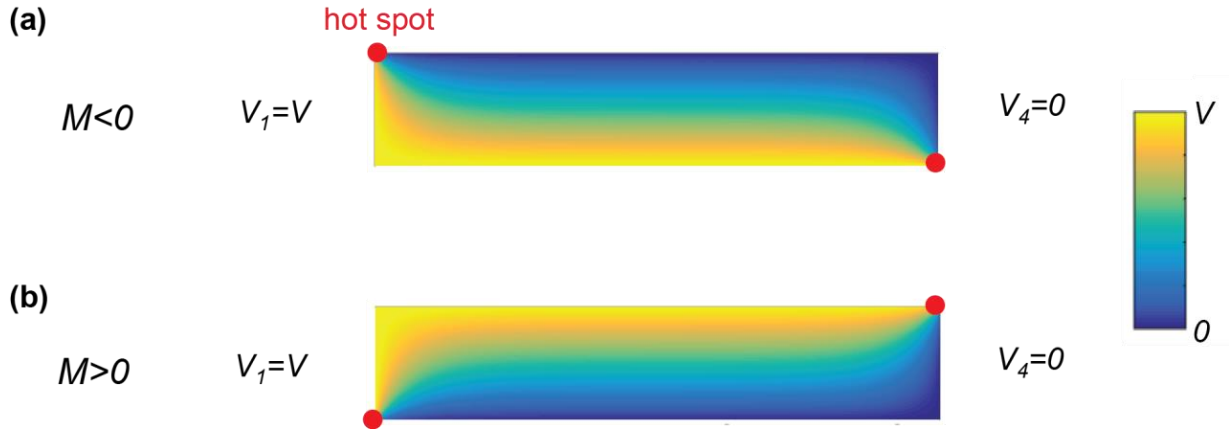

**Fig. S9. The electric potential distributions of six terminals Hall bridge in the QAH regime for *M*<0 (a) and *M*>0 (b).** The calculation assumes the tangent of Hall angle tan$\alpha$=7000, the length and the width of the samples are *L* and *W*, and *L*=5*W*.

**vii) The excitation gap estimation from the temperature dependence of the longitudinal resistance**

The zero-field longitudinal sheet resistance $\rho_{14,23}(0)$ (blue curve in **Fig. 3c** of the main text) can be fitted well with a formula below

$$\rho_{14,23}(0) = R_0 f(E_g) = \frac{R_0}{e^{\frac{E_g}{kT}} + 1} \qquad (1)$$

which has two fitting parameters $R_0$ and $E_g$, where $E_g$ can be taken as the excitation gap.

The above formula can be derived from Landauer-Buttiker theory [S3, S4], and the temperature dependence can be easily included into the transmission coefficients, as seen from the temperature dependent Landauer-Buttiker formula

$$I_p = \frac{e}{h} \int dE \sum_q T_{pq}(f_p - f_q),$$

Here $f_p = \frac{1}{e^{\beta(E-\mu_p)}+1}$ is the Fermi distribution function. For simplicity, we assume $\mu_p = E_f + eV_p$ where $V_p \ll E_f$, so that the system is in the linear response regime. Therefore,

$$f_p - f_q = f_p - f_0 - f_q + f_0 = \left(-\frac{\partial f_0}{\partial E}\right)eV_p - \left(-\frac{\partial f_0}{\partial E}\right)eV_q = \left(-\frac{\partial f_0}{\partial E}\right)e(V_p - V_q)$$

and

$$I_p = \frac{e}{h} \int dE \sum_q T_{pq}(f_p - f_q) = \frac{e^2}{h} \sum_q T_{pq}(V_p - V_q) \int dE \left(-\frac{\partial f_0}{\partial E}\right)$$

$$= \frac{e^2}{h} \sum_q T_{pq} f_0(E_g)(V_p - V_q),$$

Here the excitation gap $E_g$ can be defined as the energy difference between band bottom of non-chiral edge modes and the maximum of the bulk valence band. Therefore, one can clearly see that the temperature dependence is only included into the transmission for non-chiral modes, $T_{pq} \to T_{pq} f_0(E_g)$.

Thus, we can start from the zero temperature Landauer-Buttiker formula

$$I_i = \sum_j G_{ij}(V_i - V_j), \quad G_{ij} = \frac{e^2}{h} T_{ij}$$

where $T_{ij}$ is the transmission coefficient from electrode $j$ to electrode $i$. Here $i, j$ represent four terminals 1, 2, 3 and 4, as shown in **Fig. S10a**. We assume the terminals 2 and 3 are voltage probes, so the currents flowing into these two leads are zero $I_2 = I_3 = 0$. The explicit form of Buttiker formula should be given as

$$\begin{pmatrix} I \\ 0 \\ 0 \end{pmatrix} = \begin{pmatrix} G_{12} + G_{13} + G_{14} & -G_{12} & -G_{13} \\ -G_{21} & G_{21} + G_{23} + G_{24} & -G_{23} \\ -G_{31} & -G_{32} & G_{31} + G_{32} + G_{34} \end{pmatrix} \begin{pmatrix} V_1 \\ V_2 \\ V_3 \end{pmatrix}$$

The matrix of the transmission coefficient takes the form: $T_{ij} = \eta \delta_{i,j+1} + t_{ij}$, where $\eta$ is for the contribution of edge states and $t_{ij}$ is for bulk contribution. In addition, we assume $t_{23} = t_{32} = T_1$ and $t_{12} = t_{21} = t_{34} = t_{43} = T_2$ for simplicity.

Following the assumptions, we have

$$\begin{pmatrix} I \\ 0 \\ 0 \end{pmatrix} = \begin{pmatrix} T_2 + \eta & -T_2 & 0 \\ -(\eta + T_2) & \eta + T_2 + T_1 & -T_1 \\ 0 & -(\eta + T_1) & T_1 + \eta + T_2 \end{pmatrix} \begin{pmatrix} V_1 \\ V_2 \\ V_3 \end{pmatrix} \quad (2)$$

The equations can be simplified. From the third equation in (2), we have

$$V_3 = \frac{(\eta + T_1)V_2}{\eta + T_2 + T_1}.$$

From the second equation of (2), we have

$$V_1 = \frac{1}{\eta + T_2}((\eta + T_2 + T_1)V_2 - T_1 V_3) = \frac{V_2}{\eta + T_2}\left((\eta + T_2 + T_1) - \frac{(\eta + T_1)T_1}{\eta + T_2 + T_1}\right)$$

Substituting these equations into the first equation of (2), we have

$$I = (T_2 + \eta)V_1 - T_2 V_2 = V_2\left((\eta + T_2 + T_1) - \frac{(\eta + T_1)T_1}{\eta + T_2 + T_1}\right) - T_2 V_2$$

$$= (\eta + T_1)V_2 - \frac{(\eta + T_1)T_1}{\eta + T_2 + T_1}V_2 = (\eta + T_1)V_2\left(1 - \frac{T_1}{\eta + T_2 + T_1}\right)$$

$$= (\eta + T_1)V_2 \frac{\eta + T_2}{\eta + T_2 + T_1}.$$

Here we have omitted the coefficient $\frac{e^2}{h}$. The longitudinal resistance thus takes the form:

$$R = \frac{V_2 - V_3}{I} = \frac{V_2}{I}\left(1 - \frac{\eta + T_1}{\eta + T_1 + T_2}\right) = \frac{\eta + T_1 + T_2}{(\eta + T_1)(\eta + T_2)} \frac{T_2}{\eta + T_1 + T_2} = \frac{T_2}{(\eta + T_1)(\eta + T_2)}$$

For $T_1, T_2 \ll \eta \approx 1$, one will have

$$R = \frac{h}{e^2}\frac{T_2}{\eta^2}.$$

Now we consider the temperature dependence of $T_2$, when $T_2 \to T_2 f(E_g)$. The longitudinal resistance is thus expressed as

$$\rho_{14,23}(0) = R_0 f(E_g) = \frac{R_0}{e^{\frac{E_g}{kT}} + 1} \qquad (1)$$

where $R_0 = \frac{h}{e^2}\frac{T_2}{\eta^2}$.

Therefore, the excitation gap $E_g$ (shown in **Fig. 4o** of the main text) can estimated by the fitting of temperature dependence of $\rho_{14,23}(0)$, as shown in **Fig. S10b**. The fitting gives the parameters $R_0 = 47.7 k\Omega$ and $E_g = 50 \mu eV$. This excitation gap $50 \mu eV$ is much smaller than the gap induced by ferromagnetism [S8] and is consistent with the low observing temperature for the QAH effect.

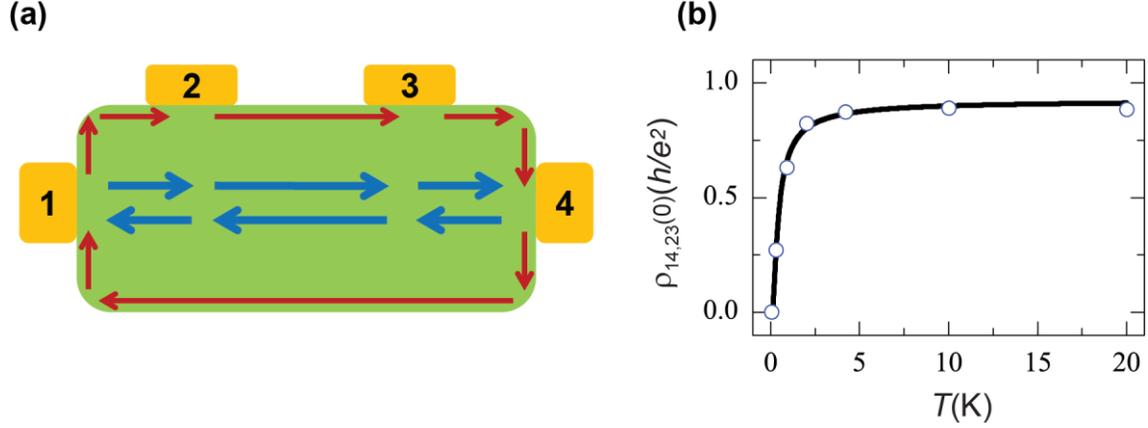

**Fig. S10. The excitation gap estimation from the temperature dependence of the longitudinal resistance.** (**a**) The schematic diagram for the theoretical calculations in (vii). The red and blue arrows indicate the chiral edge mode and nonchiral edge mode in the sample, respectively. (**b**) The theoretical fit of the temperature dependence of zero-field longitudinal sheet resistance $\rho_{14,23}(0)$ using the formula equation (1) with the parameters $R_0 = 47.7 k\Omega$ and $E_g = 50 \mu eV$.